\documentclass[12pt]{article}
\usepackage{amsmath}
\usepackage{cite}
\usepackage{graphicx}
\usepackage{amsfonts}
\usepackage{amssymb}
\csname @addtoreset\endcsname{equation}{section}
\newcommand{\eq}{\begin{equation}}
\newcommand{\feq}{\end{equation}}
\newcommand{\eqn}{\begin{eqnarray}}
\newcommand{\feqn}{\end{eqnarray}}
\newcommand{\arr}{\begin{eqnarray*}}
\newcommand{\farr}{\end{eqnarray*}}

\begin{document}
\begin{titlepage}
\begin{flushright}
CAMS/ 04-02\\
UCLA/04/TEP/27\\
hep-th/0407147\\
\end{flushright}
\vspace{.3cm}
\begin{center}
\renewcommand{\thefootnote}{\fnsymbol{footnote}}
{\Large{\bf S-brane solutions in gauged and ungauged supergravities}}
\vskip1cm
{\large\bf Michael Gutperle$^1$\footnote{email:
gutperle@physics.ucla.edu  }
and Wafic Sabra$^2$\footnote{email: ws00@aub.edu.lb}}
\renewcommand{\thefootnote}{\arabic{footnote}}
\setcounter{footnote}{0}
\vskip1cm
{\small
$^1$Department of Physics and Astronomy, UCLA, Los Angeles, CA 90095, USA\\
\vspace*{0.4cm}
$^2$ Center for Advanced Mathematical Sciences (CAMS)
and\\
Physics Department, American University of Beirut, Lebanon.\\}
\end{center}
\vskip2cm
\begin{center}
{\bf Abstract}
\end{center}
\medskip
\noindent We construct time dependent S-brane solutions in   gauged
and ungauged supergravities in various dimensions. The supergravity
solutions we find are all special cases of solutions in
gauged supergravities with
symmetric
potentials. We discuss some properties of these solutions and their relation
to topological  black holes in anti-de Sitter spaces.
\end{titlepage}

\baselineskip=20pt

\section{Introduction}

Time dependent solutions in string theory and supergravities have been of
great interest recently. Classical solutions of supergravity theories can be
the starting point to analyze questions of stability, particle creation and
singularity resolution in string theory. It is important to understand these
issues better in order to apply string theory to cosmology.

A particular class of such solutions are called S-branes (for `spacelike'
branes) \cite{Gutperle:2002ai}. S-branes describe a shell of radiation 
moving
in from infinity and forming an unstable brane which subsequently decays.
Since the brane only exists for a finite time it is localized in time and is
therefore spacelike.

A concrete realization of this idea was developed by Sen in terms of open
string tachyon condensation in \cite{Sen:2002nu,Sen:2002in,Gutperle:2003xf}.
Recently various supergravity S-brane solutions were found in
\cite{Chen:2002yq,Kruczenski:2002ap, Roy:2002ik, Ohta:2003uw, 
Burgess:2002vu,
Quevedo:2002tm, Cornalba:2002fi, Cornalba:2002nv,Deger:2002ie} (For related earlier work
see \cite{Lukas:1996ee, Lukas:1996iq, Lu:1996er, Lu:1996jk}). The ansatz for
an S-p brane supergravity solution in $d$ dimensions has $ISO(p+1)\times
SO(d-p-2,1)$ symmetry. The first factor corresponds to the symmetry of the
(spacelike) worldvolume and the second factor to the symmetry of the
transverse lightcone. In the following we will only consider S-0 branes with
one dimensional worldvolume.

Note that all of these solutions (as well as the ones we find) have
singularities. However nonsingular S-brane solutions have been obtained
recently in \cite{Jones:2004rg, Wang:2004by,Tasinato:2004dy,Lu:2004ye}.

In this note we find S-brane solutions in gauged and ungauged 
supergravities.
In section 2 we review and discuss how to obtain S-brane solutions by
analytically continuing black hole solutions. In the case of black holes in
AdS-space we point out a relation of S-brane solutions to topological AdS
black holes \cite{Liu:2003px}.

In section 3 the equations of motion for an S-brane in a theory with
nontrivial scalars and gauge fields in arbitrary dimensions are presented. 
In
section 4 we present S-brane solutions in a gauged supergravity where the
scalars lie in a coset $SL(n,R)/SO(n,R)$ and the potential is given by a
symmetric function of the scalars. The solutions are special because the
metric, scalars and gauge fields are all expressed in terms of harmonic
functions. In section 5 to 7 we find simple solutions of some particular
(un)gauged supergravities in dimensions $d=5,4$ and $7$ respectively. These
solutions can be obtained from the solution presented in section 4 by an
identification of scalar and gauge fields.

All the solutions we find have similar properties: The reality of the 
solution
imposes constraints on the charges of the solution and the `nonextremality'
parameter. In the case of gauged supergravity the S-brane solutions are
asymptotically AdS and are related to topological black holes in AdS.

\section{ Black hole solutions and S-branes in flat space and AdS}

Many S-brane and related time dependent solutions can be obtained by 
analytic
continuation of known static solutions. In this section we will review this
method using some simple examples which have been previously discussed in 
the
literature \cite{Quevedo:2002tm,Cornalba:2002fi, Cornalba:2002nv}. We start
with the metric for the four dimensional Schwarzschild black hole.
\begin{equation}
ds^{2}=-(1-{\frac{m}{r}})dt^{2}+{\frac{1}{1-{\frac{m}{r}}}}dr^{2}%
+r^{2}(d\theta^{2}+\sin^{2}\theta d\phi^{2}). \label{ssone}%
\end{equation}
A time dependent solution can be obtained from (\ref{ssone}) by replacing
\begin{equation}
r\rightarrow i\tau,\quad t\rightarrow ix,\quad m\rightarrow im,\quad
\theta\rightarrow i\theta,
\end{equation}
which gives
\begin{equation}
ds^{2}=-{\frac{1}{1-{\frac{m}{\tau}}}}d\tau^{2}+(1-{\frac{m}{\tau}}%
)dx^{2}+\tau^{2}(d\theta^{2}+\sinh^{2}\theta d\phi^{2}). \label{sstwo}%
\end{equation}
The continuation turns the metric on the two sphere $S_{2}$ into the metric 
on
$H_{2}$, the Poincar$\acute{\mbox{e}}$ plane (or Euclidean $AdS_{2}$). There
is a coordinate singularity at $t=m$ which defines a horizon. Continuing the
coordinates, one gets a static metric and a timelike curvature singularity 
at
$t=0$. The Penrose diagram of this spacetime is the one of the Schwarzschild
black hole which is rotated by $90$ degrees.

Another example is given by the Reissner-Nordstr\"{o}m solution of
Einstein-Maxwell theory
\begin{align}
ds^{2}  &  
=-(1-{\frac{2m}{r}}+{\frac{q^{2}}{r^{2}}})dt^{2}+(1-{\frac{2m}{r}%
}+{\frac{q^{2}}{r^{2}}})^{-1}dr^{2}+r^{2}(d\theta^{2}+\sin^{2}\theta d\phi
^{2}),\nonumber\\
F_{rt}  &  ={\frac{q}{r^{2}}}.
\end{align}
The analytic continuation is given by
\begin{equation}
r\rightarrow i\tau,\quad t\rightarrow ix,\quad m\rightarrow\pm im,\quad
q\rightarrow q,\quad\theta\rightarrow i\theta\label{anacon}%
\end{equation}
and produces the metric
\begin{align}
ds^{2}  &  =-(1\mp{\frac{2m}{\tau}}-{\frac{q^{2}}{\tau^{2}}})^{-1}d\tau
^{2}+(1\mp{\frac{2m}{\tau}}-{\frac{q^{2}}{\tau^{2}}})dx^{2}+\tau^{2}%
(d\theta^{2}+\sinh^{2}\theta d\phi^{2}),\nonumber\\
F_{\tau x}  &  ={\frac{q}{\tau^{2}}}.
\end{align}
An extremal black hole obeys $m=\pm q$. In this case analytic continuation
(\ref{anacon}) produces a complex field-strength and the solution obtained 
is
therefore unphysical. This indicates a generic feature that the S-brane
solutions obtained by analytic continuation are always nonsupersymmetric, as
expected from spacetimes without timelike or null Killing vectors.

The next example we discuss is a black hole in $AdS_{5}$, whose metric is
given by
\begin{equation}
ds^{2}=-(1-{\frac{m}{r^{2}}}+r^{2})dt^{2}+{\frac{1}{1-{\frac{m}{r^{2}}}+r^{2}%
}}dr^{2}+r^{2}(d\theta^{2}+\sin^{2}\theta d\Omega_{2}^{2}),
\end{equation}
where $d\Omega_{2}^{2}$ is the metric on the unit two sphere. As before one
can obtain a real time dependent solution from the above metric by 
performing
the analytic continuation \footnote{A different analytic continuation was
given in \cite{Lu:2003iv} relating the AdS BH to fluxbranes
\cite{Costa:2000nw, Gutperle:2001mb}}
\begin{equation}
r\rightarrow i\tau,\quad t\rightarrow ix,\quad\theta\rightarrow i\theta,
\label{anaconb}%
\end{equation}
which should be associated with an S-brane in AdS. The analytically 
continued
metric is
\begin{equation}
ds^{2}=(1+{\frac{m}{\tau^{2}}}-\tau^{2})dx^{2}-{\frac{1}{1+{\frac{m}{\tau^{2}%
}}-\tau^{2}}}d\tau^{2}+\tau^{2}(d\theta^{2}+\sinh^{2}\theta 
d\Omega_{2}^{2}).
\end{equation}
There are coordinate singularities at $\tau_{\pm}^{2}={\frac{1}{2}}(1\pm
\sqrt{1+4m})$. For $0>m>-1/4$ there are two horizons $\tau_{\pm}$, whereas 
for
$m>0$ there is only one horizon at $\tau_{+}$. For $\tau>\tau_{+}$ the 
metric
becomes static and one gets
\begin{equation}
ds^{2}=-(-1-{\frac{m}{\tau^{2}}}+\tau^{2})dx^{2}+{\frac{1}{-1-{\frac{m}%
{\tau^{2}}}+\tau^{2}}}d\tau^{2}+\tau^{2}dH_{3}^{2}.
\end{equation}
This metric is the `topological' black hole solution 
\cite{Birmingham:1998nr,
Mann:1996gj, Vanzo:1997gw}, i.e. a black hole in AdS with a hyperbolic (or
quotients thereof) horizon. In contrast to the asymptotically flat case, in
AdS the analytic continuation (\ref{anaconb}) of a black hole with a 
spherical
horizon does not produce a new time dependent background but instead 
produces
the inside of a black hole with a hyperbolic horizon. Since the horizon is
non-compact, this solution might have a cosmological interpretation.

\section{General Considerations}

In this section we consider theories of $d$-dimensional gravity coupled to
gauge and scalar fields described by the general Lagrangian
\begin{equation}
e^{-1}\mathcal{L}_{d}=\!\!\text{ }R-{\frac{1}{2}}\mathsf{\mathcal{M}}%
_{ij}\partial_{\mu}\phi^{i}\partial^{\mu}\phi^{j}-{\frac{1}{4}}G_{IJ}F_{\mu
\nu}^{I}F^{\mu\nu\,J}-l^{2}\mathcal{P}. \label{act}%
\end{equation}
Here $\mathcal{M}_{ij}$, $G_{IJ}$ and $\mathcal{P}$ are functions of the
scalar fields $\phi^{i}.$ The Einstein gravitational equations derived from
(\ref{act}) are given by
\begin{equation}
R_{\mu\nu}={\frac{1}{2}}\mathcal{M}_{ij}\partial_{\mu}\phi^{i}\partial_{\nu
}\phi^{j}+{\frac{1}{2}}G_{IJ}\left(  F_{\mu\lambda}^{I}F_{\nu}{}^{\lambda
\,J}-{\frac{1}{2(d-2)}}g_{\mu\nu}F_{\rho\sigma}^{I}F^{\rho\sigma\,J}\right)
+{\frac{g_{\mu\nu}}{d-2}l}^{2}\mathcal{P}. \label{ein}%
\end{equation}
The variation of (\ref{act}) with respect to the scalar and gauge fields
gives, respectively, the scalar equation of motion
\begin{equation}
{\frac{1}{\sqrt{g}}}\partial_{\mu}(\sqrt{g}g^{\mu\nu}\mathcal{M}_{ij}%
\partial_{\nu}\phi^{j})-{\frac{1}{2}}\partial_{i}\mathcal{M}_{kl}\partial
_{\mu}\phi^{k}\partial^{\mu}\phi^{l}-{\frac{1}{4}}\partial_{i}G_{JK}F_{\mu\nu
}^{J}F^{K\mu\nu}-l^{2}\partial_{i}\mathcal{P}=0
\end{equation}
and the equations of motion for the abelian gauge fields
\begin{equation}
D_{\mu}(G_{IJ}F^{\mu\nu\;J})=0.
\end{equation}
For a solution we take the following ansatz for the metric%

\begin{equation}
ds^{2}=-e^{2A(t)}dt^{2}+e^{2B(t)}dx^{2}+e^{2C(t)}ds_{d-2,k}^{2}%
\end{equation}
where $ds_{d-2,k}^{2}$ $=\bar{g}_{ab}dx^{a}dx^{b}$ ($a,b=1,..,d-2)$ and the
Ricci tensor for the metric $\bar{g}_{ab}$ is given by $\bar{R}_{ab}%
=k(d-3)\bar{g}_{ab},$ with $k=1,-1,0$. We also take $A_{x}^{I}(t)$ to be the
only non-vanishing component of the vector potentials. The non-vanishing
components of the Ricci tensor are then given by%

\begin{align}
R_{xx}  &  =e^{2B-2A}(B^{\prime\prime}+B^{\prime2}-A^{\prime}B^{\prime
}+(d-2)B^{\prime}C^{\prime}),\nonumber\\
R_{tt}  &  =-(B^{\prime\prime}+B^{\prime2}-A^{\prime}B^{\prime})-(d-2)\left(
C^{\prime\prime}+C^{\prime2}-A^{\prime}C^{\prime}\right)  ,\nonumber\\
R_{ab}  &  =\left[  
e^{2C-2A}(C^{\prime\prime}-A^{\prime}C^{\prime}+B^{\prime
}C^{\prime}+(d-2)C^{\prime^{2}})+k(d-3)\right]  \bar{g}_{ab}.
\end{align}
The Einstein equations of motion (\ref{ein}) give for our ansatz
\begin{align}
B^{\prime\prime}+B^{\prime}\left(  B^{\prime}-A^{\prime}+(d-2)C^{\prime
}\right)   &  =\chi,\nonumber\\
-(B^{\prime\prime}+B^{\prime2}-A^{\prime}B^{\prime})-(d-2)\left(
C^{\prime\prime}+C^{\prime2}-A^{\prime}C^{\prime}\right)   &  ={\frac{1}{2}%
}\mathcal{M}_{ij}\partial_{t}\phi^{i}\partial_{t}\phi^{j}-\chi,\nonumber\\
C^{\prime\prime}+C^{\prime}\left(  -A^{\prime}+B^{\prime}+(d-2)C^{\prime
}\right)  +ke^{2A-2C}(d-3)  &  =\chi+\frac{\mathcal{F}}{2}e^{-2B}, 
\label{gem}%
\end{align}
where%
\begin{equation}
\chi=\frac{1}{2(d-2)}\left(  -(d-3)e^{-2B}\mathcal{F}+2l^{2}{e^{2A}%
\mathcal{P}}\right)
\end{equation}
and $\mathcal{F}=G_{IJ}F_{tx}^{I}F_{tx}^{J}$. The equations of motion 
simplify
with the following relations
\begin{align}
e^{2B}  &  =fe^{-2(d-3)V},\quad e^{2A}=\frac{e^{2V}}{f},\quad e^{2C}%
=t^{2}e^{2V},\nonumber\\
\text{ }f  &  =-k+\frac{\mu}{t^{d-3}}-l^{2}t^{2}e^{2(d-2)V},
\end{align}
and one obtains from (\ref{gem}) that the scalar fields must satisfy
\begin{equation}
{\frac{1}{2}}\mathcal{M}_{ij}\partial_{t}\phi^{i}\partial_{t}\phi^{j}=-\left(
d-2\right)  \left[  \frac{1}{t}V^{\prime}\left(  d-2\right)  +V^{\prime}{}%
^{2}\left(  d-3\right)  +V^{\prime\prime}\right]  . \label{gravs}%
\end{equation}
We also obtain from (\ref{gem}) the following relation for the gauge fields%

\begin{equation}
\mathcal{F}=-2(d-2)e^{-2(d-3)V}\left[  k\left(  V^{\prime\prime}%
+\frac{V^{\prime}}{t}\left(  d-2\right)  \right)  -\frac{\mu}{t^{d-3}}\left(
V^{\prime\prime}+\frac{1}{t}V^{\prime}\right)  \right]  . \label{gf}%
\end{equation}
Note that the $l^{2}$ drops out in the expression for the gauge fields and
that the scalar fields are independent of the parameters $\mu$ and $k$.

As a special case, we consider theories with trivial scalars, i.e., Einstein
Maxwell theory with a potential (negative cosmological constant)
$\mathcal{P}=-(d-2)(d-3)$. In these cases, we obtain from (\ref{gravs})%

\begin{equation}
V^{\prime\prime}+\frac{V^{\prime}}{t}\left(  d-2\right)  +V^{\prime}{}%
^{2}\left(  d-3\right)  =0.
\end{equation}
The above equation can be solved by:
\begin{equation}
e^{V}=\left(  1+\frac{q}{t^{d-3}}\right)  ^{\frac{1}{d-3}}.
\end{equation}
When substituted in (\ref{gf}), this solution gives%

\begin{equation}
\mathcal{F}=2(d-2)\left(  d-3\right)  q\frac{\left(  kq+\mu\right)  }%
{t^{2d-4}\left(  1+\frac{q}{t^{d-3}}\right)  ^{4}}.
\end{equation}
Note that reversing the sign of $l^{2}$ gives solutions in $d$-dimensional 
de
Sitter Einstein Maxwell theories. These general solutions were considered in
\cite{Liu:2003px}.

For theories with non-trivial scalars, one has to solve a more complicated 
set
of equations. A natural ansatz is to express $V$ and the gauge fields in 
terms
of harmonic functions. Equation (\ref{gravs}) can then be used to determine
the scalar fields in terms of the harmonic functions. However for a general
potential the scalar fields will not solve the other equations of motion. In
what follows we will analyze special cases of gauged supergravity theories 
in
which the harmonic ansatz works.

\section{Symmetric potentials}

In this section we consider solutions of a gauged supergravity where the
scalars parameterize the subspace $SL(N,R)/SO(N,R)$ of the coset manifold of
maximal gauged supergravity \footnote{Domain wall solutions in these 
theories
were discussed in \cite{Cvetic:1999xx}.}. The Lagrangian in dimension $d$ is
given by%

\begin{equation}
e^{-1}\,\mathcal{L}_{d}=R-{{\frac{{1}}{{2}}}}(\partial\vec{\varphi}%
)^{2}-{\frac{1}{4}}G_{IJ}F_{\mu\nu}^{I}F^{\mu\nu\,J}-l^{2}\mathcal{P}\,,
\label{ddlag}%
\end{equation}
where the potential $\mathcal{P}$ is a symmetric potential given by
\begin{equation}
\mathcal{P}=-\frac{(d-3)^{2}}{8}\,\left(  (\sum_{I=1}^{N}X_{I})^{2}%
-2\sum_{I=1}^{N}X_{I}^{2}\right)  \,. \label{ddpot}%
\end{equation}
We consider theories with gauge kinetic term given by
\begin{equation}
G_{IJ}={\frac{1}{(X_{I})^{2}}}\delta_{IJ}.
\end{equation}
The $N$ scalars $X_{I}$ are subject to the constraint
\begin{equation}
\prod_{I=1}^{N}X_{I}=1\,. \label{prodcon}%
\end{equation}
The constrained scalars $X_{I}$ are parameterized in terms of $(N-1)$
independent dilatonic scalars $\vec{\varphi}$ as follows
\begin{equation}
X_{I}=e^{-{\frac{1}{2}}\vec{b}_{I}\cdot\vec{\varphi}}\,, \label{ddxdef}%
\end{equation}
where the $\vec{b}_{I}$ are the weight vectors of the fundamental
representation of $SL(N,R)$, satisfying
\begin{equation}
\vec{b}_{I}\cdot\vec{b}_{J}=8\delta_{IJ}-{\frac{8}{N}}\,,\qquad\sum_{I}\vec
{b}_{I}=0\,,\qquad(\vec{u}\cdot\vec{b}_{I})\,\vec{b}_{I}=8\vec{u}\,.
\label{dotprod}%
\end{equation}
The vector $\vec{u}$ is an arbitrary $N$-vector. The above relations allow
$\vec{\varphi}$ to be determined in terms of $X_{I}$
\begin{equation}
\vec{\varphi}=-{{\frac{{1}}{{4}}}}\sum_{I}\vec{b}_{I}\,\log X_{I}\,.
\end{equation}
We take the same ansatz for the metric as in the previous section
\begin{equation}
ds^{2}=fe^{-2(d-3)V}dx^{2}+e^{2V}\left(  -\frac{dt^{2}}{f}+t^{2}ds_{d-2,k}%
^{2}\right)  .
\end{equation}
We find that the Einstein equations of motion admit solutions given by
\begin{align}
e^{2V}  &  =\prod\left(  H_{I}\right)  ^{\frac{1}{2\left(  d-2\right)  }%
},\text{ \ \ \ }f=-k+\frac{\mu}{t^{d-3}}-l^{2}t^{2}e^{2(d-2)V},\nonumber\\
X_{I}  &  =\frac{1}{H_{I}}\prod_{J}\left(  H_{J}\right)  ^{\frac{1}{4}\left(
\frac{d-3}{d-2}\right)  }=\frac{1}{H_{I}}e^{(d-3)V},\nonumber\\
F_{xt}^{I}  &  =\frac{\left(  d-3\right)  }{H_{I}^{2}t^{d-2}}\sqrt{\left(
kq_{I}^{2}+\mu q_{I}\right)  },\nonumber\\
H_{I}  &  =1+{\frac{q_{I}}{t^{d-3}}}.
\end{align}
provided the following relation holds
\begin{equation}
\vec{b}_{I}.\vec{b}_{J}=8\delta_{IJ}-2\frac{d-3}{d-2}.
\end{equation}
It follows from the first equation in (\ref{dotprod}) that $N=4(d-2)/(d-3)$.
Since $N$ and $d$ have to be integers, this relation holds for $d=4,5,7$,
where $N=8,6,5$ respectively.

\bigskip We now turn to the scalar equation of motion which for our ansatz
reads
\begin{equation}
-{\frac{1}{4}}\partial_{\vec{\phi}}G_{JK}F_{\mu\nu}^{J}F^{K\mu\nu}+{\frac
{1}{\sqrt{g}}}\partial_{t}(\sqrt{g}g^{tt}\partial_{t}\vec{\phi})-l^{2}%
\partial_{\vec{\phi}}\mathcal{P}=0.
\end{equation}
Using the fact that $\sum_{I}\vec{b}_{I}=0,$ as well as $\frac{\partial 
X_{I}%
}{\partial\vec{\phi}}=-\frac{1}{2}\vec{b}_{I}X_{I},$ it is easy to show that
the scalar equation of motion is satisfied for the scalar potential given in
(\ref{ddpot}).

\section{Five dimensional $N=2$ gauged supergravity}

Here we will consider the five dimensional supergravity theory obtained from
dimensionally reducing type-IIB supergravity on $S^{5}.$ This theory has 
three
abelian vector multiplets (including the graviphoton). The scalars
$X_{I}=S,T,U$ of this theory satisfy the constraint $X_{1}X_{2}X_{3}=1$ and
thus the theory has two independent scalar fields. Taking $X_{1},X_{2}$ as 
the
independent variables, the potential for this theory is given by
\begin{equation}
\mathcal{P}=-4\Big({\frac{1}{X_{1}}}+{\frac{1}{X_{2}}}+X_{1}X_{2}\Big).
\end{equation}
The action can be written in the form (\ref{act}) with%
\[
\mathcal{M}_{ij}=\left(
\begin{array}
[c]{cc}%
{\frac{2}{X_{1}^{2}}} & {\frac{1}{X_{1}X_{2}}}\\
{\frac{1}{X_{1}X_{2}}} & {\frac{2}{X_{2}^{2}}}%
\end{array}
\right)  ,\quad\quad G_{IJ}=\left(
\begin{array}
[c]{ccc}%
{\frac{1}{X_{1}^{2}}} & 0 & 0\\
0 & {\frac{1}{X_{2}^{2}}} & 0\\
0 & 0 & X_{1}^{2}X_{2}^{2}%
\end{array}
\right)  .
\]
Here we have ignored a Chern-Simons term which is not relevant for our
solutions. Note that if we write $X_{I}=e^{-\frac{1}{2}\left(  \vec{a}%
_{I.}\vec{\varphi}\right)  }$ with
\begin{equation}
\ \vec{a}_{1}=\left(  \frac{2}{\sqrt{6}},\sqrt{2}\right)  ,\text{ \ \ }\vec
{a}_{2}=\left(  \frac{2}{\sqrt{6}},-\sqrt{2}\right)  ,\text{ \ \ }\vec{a}%
_{3}=\left(  -\frac{4}{\sqrt{6}},0\right)  ,
\end{equation}
then the kinetic term for the scalar fields takes the canonical form
$-{\frac{1}{2}}\left(  \partial\vec{\phi}\right)  ^{2}.$

The ansatz for S-brane metric is
\begin{equation}
ds^{2}=e^{-4V(t)}f(t)dx^{2}-{\frac{e^{2V}}{f(t)}}dt^{2}+e^{2V}t^{2}%
(d\theta^{2}+\sinh^{2}\theta d\Omega_{2}^{2}),\label{fivedm}%
\end{equation}
where $e^{6V}$ and $f$ are given by
\begin{equation}
e^{6V}=h_{1}(t)h_{2}(t)h_{3}(t),\text{ \ \ \ \ \ }f(t)=1+{\frac{\mu}{t^{2}}%
}-l^{2}t^{2}e^{6V}.\label{fivedf}%
\end{equation}
and $h_{I}$ , $I=1,2,3,$ is a harmonic function given by
\begin{equation}
h_{I}(t)=1+{\frac{q_{I}}{t^{2}}}.
\end{equation}
The scalars and the gauge field-strength can be expressed as
\begin{equation}
X_{I}={\frac{e^{2V(t)}}{h_{I}(t)}},\text{ \ \ }F_{tx}^{I}={\frac{2}%
{h_{I}(t)^{2}}}{\frac{{\tilde{q}}_{J}}{t^{3}}.}%
\end{equation}
All equations of motion will be satisfied if the following relation between
$q_{I},\tilde{q}_{i}$ holds
\begin{equation}
(\tilde{q}_{I})^{2}=-q_{I}^{2}+\mu q_{I},\label{chargerel}%
\end{equation}
which can be represented as
\begin{equation}
q_{I}=\mu\sin^{2}\beta_{I},\quad\tilde{q}_{I}=\mu\sin\beta_{I}\cos\beta_{I}.
\end{equation}
Note that (\ref{chargerel}) implies that it is impossible to have an 
extremal
solution (where $\mu=0$) with real $\tilde{q}_{I}$, i.e. real gauge fields.

The zeros of the function $f(t)$ given in (\ref{fivedf}) determine the
location of the horizon. Continuation past the horizon produces a static
metric, which can be identified with a topological black hole solution of 
the
gauged supergravity. This is similar to the case in the simple five
dimensional AdS black hole discussed in section 2.

Setting the coupling $l=0$ in the action gives an ungauged $N=2$ 
supergravity.
The S-brane solution we have found in the gauged supergravity is also a
solution in this limit, giving an S-brane in asymptotically flat space. This
solution generalizes the S-brane related to the RN black hole found in
\cite{Gutperle:2002ai,Quevedo:2002tm,Cornalba:2002fi}. The time dependent
metric (\ref{fivedm}) has a horizon at $t=0$. Continuing the coordinates 
past
the horizon produces a static spacetime with a timelike curvature 
singularity.
The causal structure of the spacetime is very similar to the metric
(\ref{sstwo}).

The solution can be related by analytical continuation to the black hole
solution found in \cite{Behrndt:1998jd}
\begin{equation}
\tau\rightarrow ir,\quad x\rightarrow it,\quad\tilde{q}_{I}\rightarrow
i\tilde{q}_{I},\quad\theta\rightarrow i\theta.\label{anaconc}%
\end{equation}
The continued charges satisfy the condition
\begin{equation}
\tilde{q}_{I}^{2}=q_{I}^{2}+\mu q_{I}\label{fivecon}%
\end{equation}
which can be solved by
\begin{equation}
q_{I}=\mu\sinh^{2}\beta_{I},\quad\tilde{q}_{I}=\mu\sinh\beta_{I}\cosh\beta
_{I}.
\end{equation}
Note that for the black hole solutions, it is possible to obtain extremal
solutions by setting $\mu=0$ while maintaining the reality of the gauge 
field.

\section{A gauged supergravity in four dimensions}

The Kaluza Klein reduction of eleven dimensional supergravity on $S^{7}$ 
gives
rise to $N=8$, $d=4$ gauged supergravity with gauge group $SO(8)$. There
exists an abelian truncation with four gauge fields $A_{I}$, and four 
scalars
$X_{I}$ satisfying the constraint $X_{1}X_{2}X_{3}X_{4}=1$
\cite{Cvetic:1999xp,Duff:1999gh,Sabra:1999ux}. If we choose the independent
scalars to be $X_{1},X_{2}$ and $X_{3}$, the potential of the gauged theory
will take the form
\begin{equation}
\mathcal{P}=-\left(  X_{1}X_{2}+X_{1}X_{3}+X_{2}X_{3}+{\frac{1}{X_{1}X_{2}}%
}+{\frac{1}{X_{1}X_{3}}}+{\frac{1}{X_{2}X_{3}}}\right)  .
\end{equation}
The metrics $\mathcal{M}_{ij}$ and $G_{IJ}$ are given by
\begin{equation}
\mathcal{M}_{ij}=\left(
\begin{array}
[c]{ccc}%
{\frac{2}{X_{1}^{2}}} & {\frac{1}{X_{1}X_{2}}} & {\frac{1}{X_{1}X_{3}}}\\
{\frac{1}{X_{1}X_{2}}} & {\frac{2}{X_{2}^{2}}} & {\frac{1}{X_{2}X_{3}}}\\
{\frac{1}{X_{1}X_{3}}} & {\frac{1}{X_{2}X_{3}}} & {\frac{2}{X_{3}^{2}}}%
\end{array}
\right)  ,\quad G_{IJ}=\left(
\begin{array}
[c]{cccc}%
{\frac{1}{X_{1}^{2}}} & 0 & 0 & 0\\
0 & {\frac{1}{X_{2}^{2}}} & 0 & 0\\
0 & 0 & {\frac{1}{X_{3}^{2}}} & 0\\
0 & 0 & 0 & X_{1}^{2}X_{2}^{2}X_{3}^{2}%
\end{array}
\right)  .
\end{equation}
Note that if we write $X_{I}=e^{-\frac{1}{2}\left(  \vec{a}_{I}.\vec{\phi
}\right)  }$ with
\[
\vec{a}_{1}=\left(  1,1,1\right)  ,\ \ \ \vec{a}_{2}=\left(  1,-1,-1\right)
\ ,\vec{a}_{3}=\left(  -1,1,-1\right)  ,\ \ \vec{a}_{4}=\left(
-1,-1,1\right)  ,
\]
then the kinetic term of the scalar fields takes the canonical form.

The ansatz for the S-brane metric is
\begin{equation}
ds^{2}=e^{-2V(t)}f(t)dx^{2}+e^{2V}\left(  -{\frac{1}{f(t)}}dt^{2}%
+t^{2}(d\theta^{2}+\sinh^{2}\theta d\phi^{2})\right)  ,
\end{equation}
where
\begin{equation}
f=1+{\frac{\mu}{t}}-l^{2}t^{2}e^{4V},\quad e^{4V(t)}=h_{1}(t)h_{2}%
(t)h_{3}(t)h_{4}(t).
\end{equation}
$h_{I}$ is a harmonic function given by
\begin{equation}
h_{I}(t)=1+{\frac{q_{I}}{t}}.
\end{equation}
The scalars and the gauge field-strength can be expressed as
\begin{equation}
X_{I}={\frac{e^{V}}{h_{I}}},\quad F_{tx}^{I}={\frac{{\tilde{q}}_{I}}{h_{I}%
^{2}(t)t^{2}}}.
\end{equation}
The equations of motion will be satisfied if the following relation holds%
\begin{equation}
(\tilde{q}_{I})^{2}=-q_{I}^{2}+\mu q_{I}.\label{fourcon}%
\end{equation}

\section{A gauged supergravity in seven dimension}

The Kaluza Klein reduction of eleven dimensional supergravity on $S^{4}$ 
gives
rise to a $d=7$ gauged supergravity with gauge group $SO(5)$
\cite{Pernici:xx,Townsend:1983kk}. It is possible to truncate the theory to 
an
abelian $U(1)^{2}$ subgroup, with two gauge fields $A_{I},I=1,2$ and two
scalars $X_{i},$ $i=1,2$.

\bigskip The metrics $\mathcal{M}_{ij}$ and $G_{IJ}$ and the potential are
given by$\ $
\begin{align}
\mathcal{M}_{ij} &  =\left(
\begin{array}
[c]{cc}%
{\frac{3}{X_{1}^{2}}} & {\frac{2}{X_{1}X_{2}}}\\
{\frac{2}{X_{1}X_{2}}} & {\frac{3}{X_{2}^{2}}}%
\end{array}
\right)  ,\quad G_{IJ}=\left(
\begin{array}
[c]{cc}%
{\frac{1}{X_{1}^{2}}} & 0\\
0 & {\frac{1}{X_{2}^{2}}}%
\end{array}
\right)  ,\nonumber\\
\mathcal{P} &  =-16X_{1}X_{2}-8\left(  \frac{1}{X_{1}X_{2}^{2}}+\frac{1}%
{X_{2}X_{1}^{2}}\right)  +\frac{2}{\left(  X_{1}X_{2}\right)  ^{4}}.
\end{align}
Like in four and five dimensions, the kinetic term for the scalar fields 
take
the canonical form if we write $X_{i}=e^{-\frac{1}{2}\left(  \vec{a}_{i}%
.\vec{\phi}\right)  }$ with $\vec{a}_{1}=\left(  \sqrt{2},\sqrt{\frac{2}{5}%
}\right)  ,\ \ \vec{a}_{2}=\left(  -\sqrt{2},\sqrt{\frac{2}{5}}\right)  .$

The ansatz for the S-brane solution is
\begin{align}
ds^{2} &  =e^{-8V(t)}f(t)dx^{2}+e^{2V}\left(  -{\frac{1}{f(t)}}dt^{2}%
+t^{2}(d\theta^{2}+\sinh^{2}\theta d\Omega_{4}^{2})\right)  ,\nonumber\\
f(t) &  =1+{\frac{\mu}{t^{4}}}-l^{2}t^{2}e^{10V},\quad e^{10V(t)}%
={(1+\frac{q_{1}}{t^{4}})(1+\frac{q_{2}}{t^{4}})},\nonumber\\
X_{1} &  ={e^{4V}(1+\frac{q_{1}}{t^{4}})}^{-1}{,}\text{ \ \ }X_{2}%
={e^{4V}(1+\frac{q_{2}}{t^{4}})}^{-1},\nonumber\\
F_{tx}^{1} &  =4{\frac{{\tilde{q}}_{1}}{t^{5}}(1+\frac{q_{1}}{t^{4}})}%
^{-2},\text{ \ }F_{tx}^{2}=4{\frac{{\tilde{q}}_{2}}{t^{5}}(1+\frac{q_{2}%
}{t^{4}})}^{-2}.
\end{align}
The equations of motion will be satisfied if the following relation holds%
\begin{equation}
(\tilde{q}_{I})^{2}=-q_{I}^{2}+\mu q_{I},\quad\quad I=1,2.\label{sevencon}%
\end{equation}
Note that the condition on the charges in this and the previous section is
exactly the same as in five dimensions (\ref{fivecon}). The solutions share
the same structure and properties: For example, an analytic continuation
(\ref{anaconc}) relates the S-brane solutions to the black hole solutions
found in \cite{Cvetic:1999xp}.

\section{Discussion}

In this paper we have considered S-brane solutions of gauged and ungauged
supergravity theories in various dimensions. The solutions have a close
relationship with black holes and can be obtained by an analytic 
continuation
from static black hole solutions. This analytic continuation modifies the
charge parameters and the relation (\ref{chargerel}) implies that, for the
S-brane solutions, there exists no `non-extremal' limit which keeps the 
gauge
fields in the solution real. This is in agreement with the expectation that
S-brane solutions cannot be supersymmetric since they do not posses timelike
or null Killing vectors. \footnote{However, we note that upon the 
replacement
of $l$ by $il,$ our solutions become relevant to the theories of de Sitter
supergravity \cite{Liu:2003qa} which were obtained from the reduction of 
IIB*
and M* theory\cite{Hull:1998vg}. Note also that in these models, time
dependent `extremal' solutions are possible.}

In the bosonic action of the gauged supergravities the only term which is
produced by the gauging is the scalar potential. In the limit 
$l\rightarrow0$
the gauged supergravity goes over to an ungauged supergravity. In this limit
the solutions which we presented go over to S-branes in asymptotically flat
space. They generalize known S-brane solutions \cite{Burgess:2002vu} since
they have both nontrivial gauge and scalar fields. The solutions have a
horizon at $t=0$ which separates the time dependent S-brane spacetime from a
static spacetime. In the static part of the spacetime there is a timelike
curvature singularity. In \cite{Burgess:2002vu} the timelike singularity in
the S-brane spacetime was interpreted as a negative tension object. The
supergravities we have considered have a natural interpretation as
compactifications of string theory or M theory. It would be interesting to
check whether the singularity can be identified with an orientifold plane in
the full string theory. It might also be interesting to investigate the
possibility of nonsingular S-brane solutions by constructing di-hole 
solutions
\cite{Emparan:1999au} in the supergravity theories we have considered and
analytically continuing them along the lines of \cite{Jones:2004rg}.

The apparent similarities for the three S-brane solutions we have presented
for the gauged supergravities in $d=4,5,7$ lie in the fact that they can be
obtained from the general solution of the gauged supergravity with symmetric
potentials presented in section 4. This can be done by suitable 
identification
of gauge and scalar fields. Such an identification has been discussed in
\cite{Cvetic:1999xx} in the study of domain wall solutions. It would be very
interesting to generalize the solutions for the symmetric scalar potential 
to
more general gauged supergravity theories. It is expected that the simple
harmonic ansatz will not work in the general cases and that the solutions 
will
be more complicated. It would be interesting to investigate this further and
in particular to find a criterion for when a harmonic ansatz will work.

The S-brane solutions in gauged supergravities we have found possess the
property that continuation beyond the horizon relates them to topological
black hole solutions in AdS. In this respect it might be interesting to 
study
the AdS S-brane solutions from the point of view of a dual CFT. It would 
also
be interesting to lift the solutions to ten and eleven dimensions
\cite{Cvetic:1999xp} and study their properties. We leave these questions 
for
future work. \medskip

\noindent\textbf{Acknowledgments}

\medskip

\noindent MG would like to thank H.~Lu and C.~Pope at TAMU for hospitality 
and
useful conversations at initial stages of this work. The work of MG is
supported in part by NSF grant 0245096 and the work of WS is supported in 
part
by NSF grant PHY-0313416. Any opinions, findings and conclusions expressed 
in
this material are those of the authors and do not necessarily reflect the
views of the National Science Foundation.

\end{document}